\documentstyle[prc,twocolumn,aps,epsf,eqsecnum,ifthen,epsfig]{revtex}

\newcommand{\nc}{\newcommand}
\nc{\pt}{p_{\rm T}}
\nc{\se}{\section}
\nc{\suse}{\subsection}
\nc{\beq}[1]{\begin{equation}\label{#1}}
\nc{\eeq}{\end{equation}}
\nc{\bea}[1]{\begin{eqnarray}\label{#1}}
\nc{\eea}{\end{eqnarray}}
\nc{\bce}{\begin{center}}
\nc{\ece}{\end{center}}
\nc{\bit}{\begin{itemize}}
\nc{\eit}{\end{itemize}}
\nc{\bmp}{\begin{minipage}}
\nc{\emp}{\end{minipage}}

\nc{\la}{\langle}       
\nc{\lla}{\left \langle}
\nc{\ra}{\rangle}       
\nc{\rra}{\right \rangle}

\begin{document}

\twocolumn[\hsize\textwidth\columnwidth\hsize\csname @twocolumnfalse\endcsname
\title{Momentum anisotropies in the quark coalescence model}   
\author{Peter F.~Kolb}
\address{Physik Department, Technische Universit\"at M\"unchen,
D-85747 Garching, Germany  }
\author{Lie-Wen Chen, Vincenzo Greco, and Che Ming Ko}
\address{Cyclotron Institute and Physics Department, Texas A\&M
University, College Station, Texas 77843, USA}

\today

\maketitle


\begin{abstract}
Based on the quark coalescence model, we derive relations among the 
momentum anisotropies of mesons and baryons in relativistic heavy 
ion collisions from a given, but arbitrary azimuthal distribution 
for the partons. Besides the familiar even Fourier coefficients such 
as the elliptic flow, we also pay attention to odd Fourier 
coefficients such as the directed flow, which has been observed 
at finite rapidity even at RHIC energies.

\vspace{0.2cm}
\end{abstract}
\vspace{0.2in}
]
\begin{narrowtext}
\newpage

\section{Introduction}  
\label{sec_intro}
The systematic analysis of anisotropies of particle momentum 
distributions in the plane perpendicular to the beam direction 
in non-central collisions of heavy ions offers the possibility to 
understand both the dynamics of the collisions and the properties 
of the initially produced hot and dense matter \cite{reisdorf}. 
In the low transverse momentum region, this momentum anisotropy 
is a result of the hydrodynamic expansion that is 
driven by anisotropic pressure gradients in the system 
\cite{Ollitrault92,KSH00}. Particles of intermediate to 
high energies do not follow this collective dynamics, but instead 
radiate more or less energy, depending on how much matter they
traverse on their way out of the fireball. As a result, these particles
also acquire a strong azimuthal anisotropy in their momentum 
distributions \cite{Wang01}. For heavy-ion 
collisions at the Relativistic Heavy Ion Collider (RHIC), it has 
been shown in hydrodynamical models \cite{tean,kolb,huov,KH03} that 
the lower-order momentum anisotropy, such as the elliptic flow, 
is sensitive to the equation of state of the quark-gluon plasma 
formed during the initial stage of the collision as well as the 
transition region to hadronic matter in the later stages. In transport 
models \cite{zhang,moln1,lin1}, the hadron elliptic flow is shown 
to depend on the parton scattering cross sections in the initial 
partonic matter. For hadrons with high transverse momentum, the 
elliptic anisotropy can further provide information on the energy density 
of the initial hot matter \cite{gyulv2}. 

Although the hadron anisotropies become smaller with increasing 
order, they seem to be sensitive to the initial collision dynamics as well
\cite{KSH99,Teaney99,Kolb03}. Recent experimental 
results from the \textrm{STAR} Collaboration \cite{STAR03} have 
demonstrated that some of the higher-order harmonics are 
still measurable. The experimental data further indicate that 
there is a scaling relation among the hadronic anisotropy coefficients  
$v_n$, i.e., 
$v_n(p_T)\sim 1.2 v_2^{n/2}(p_T)$. To gain insight in the origin of this 
scaling relation, we relate in the present paper the hadron anisotropies 
to those of quarks using the quark coalescence model. In particular, 
a naive quark coalescence model \cite{MV03,lin2}, which ignores the 
momentum distribution of quarks inside hadrons, will be used. 
Recently, this model has been shown to give a natural explanation for 
another scaling law observed in experiments: the momentum dependence  
of the elliptic anisotropy of identified particles (with the exception 
of pions) becomes similar if both anisotropy and momentum are divided 
by the number of constituent quarks in the hadron, i.e., two for 
mesons and three for baryons. In Ref.\cite{GK04}, it was shown 
that the deviation of the pion elliptic flow from the scaling law 
could be explained by taking into account the effect of resonance
decays as well as the momentum distribution of quarks in hadrons 
as in the more realistic quark coalescence model
\cite{FMNB03,GKL03,HY03}.

\section{Anisotropies in parton and hadron spectra}
\label{partonsandhadrons}

The azimuthal distribution of the particle transverse momentum spectra 
(or the probability of having a particle with a given transverse 
momentum $p_T$ and azimuthal angle $\varphi$) can in general be expressed
in terms of the Fourier series
\bea{equ:partondistribution}
f(\varphi, p_T) 
&\propto&
1 + 2 \sum_{n=1}^{\infty} v_n(p_T) \, \cos n \varphi \,,
\eea
where $v_n$ denotes the $n$-th order momentum anisotropy. In the 
above, we have chosen the beam direction for the $z$-axis and the 
impact parameter for the $x$-axis.  Sine terms which are in general 
possible in a Fourier expansion always vanish in these coordinates 
because of the reflection symmetry with respect to the reaction 
plane (the $(x,z)$ plane). For collisions of equal nuclei, the odd 
Fourier coefficients vanish at mid-rapidity due to the additional 
symmetry with respect to the $y$-axis. This symmetry is lost 
at forward and backward rapidities as a result of spectator matter 
from the initial nuclei, which eventually leads to a tilt of the 
reaction zone out of the transverse plane. Collisions of non-identical 
nuclei such as S+Au break this symmetry even at mid-rapidity. It is 
therefore of great interest to keep the odd coefficients in our 
general considerations.

In the simplest coalescence model \cite{MV03}, mesons of transverse momentum 
$p_T$ are formed by quarks and antiquarks of half the transverse momentum $p_T/2$ 
traveling in the same direction. Since the azimuthal distribution function 
Eq.(\ref{equ:partondistribution}), after normalizing appropriately,
can be interpreted as a probability distribution in the azimuthal angle,
the distribution function for mesons is therefore given by the 
square of that for partons, and it can also be expressed in terms of 
a Fourier series, i.e., 
\bea{equ:Fisfsquared} 
F(\varphi)  &\propto& f(\varphi)^2 
\propto 1 + 2 \sum_{n=1}^{\infty} V_n \cos n \varphi  \,, 
\eea
where $f(\varphi)$ given by Eq.(\ref{equ:partondistribution}) is
taken to be the azimuthal distribution of parton transverse momentum
spectra, and $V_n$ denotes the $n$-th order meson anisotropy. 
Following the calculations outlined in the Appendix, we arrive at 
the following expression for $V_n$ at momentum $p_T$ in terms of 
the parton anisotropy $v_n$ at momentum $p_T/2$:
\beq{equ:Vnsolution}
V_n = \frac{1}{N} \left( 2 \, v_n + \sum_{i=1}^{n-1} v_i v_{n-i} 
+ 2 \sum_{i=1}^{\infty} v_i v_{n+i} \right)\,,
\eeq
with
\beq{equ:Vnnormalization}
N = 1+ 2 \sum_{i=1}^\infty v_i^2\;.
\eeq

Similarly, we obtain the Fourier spectra of baryons by taking the 
azimuthal distribution of the parton transverse momentum spectra 
to the third power, i.e., 
\bea{equ:Fisfscubed} 
{\tilde F}(\varphi)&\propto& f(\varphi)^3 
\propto 1 + 2 \sum_{n=1}^{\infty} {\tilde V}_n \cos n \varphi  \,.
\eea
The calculations given in the Appendix then lead to a somewhat 
more complicated expression for the $n$-th order baryon anisotropy
$\tilde V_n$, 
\bea{equ:Vntildesolution}
{\tilde V}_n = \frac{1}{{\tilde N}}   &   & 
\left(3 \, v_n + 3 \sum_{i=1}^{n-1} v_i v_{n-i} + 6 
\sum_{i=1}^{\infty} v_i v_{n+i} \right. \nonumber \\
& &          +  3 \sum_{i=1}^\infty \sum_{j=1}^\infty v_i v_j v_{n+i+j}  
+ 3 \sum_{i=1}^{\infty} \sum_{j=1}^{n+i-1} v_i v_j v_{n+i-j} \nonumber \\
&  & \left.   +\sum_{i=1}^{n-1} \sum_{j=1}^{n-i-1} v_i v_j v_{n-(i+j)} 
\right) \,,                   
\eea
with
\beq{equ:Vnnormalization1}
{\tilde N} = 1+ 6 \sum_{i=1}^\infty v_i^2 + 6 \sum_{i=1}^\infty 
\sum_{j=1}^\infty v_i v_j v_{i+j}\,.
\eeq
In the above, the parton anisotropy $v_n$'s on the right hand
side of the equation are determined at a momentum that is 1/3 of 
the baryon momentum at which the baryon anisotropy $\tilde V_n$'s 
on the left side of the equation are evaluated.

Dominant and clearly visible in the expressions above is the linear connection 
between the $n$-th harmonic of the partonic sector and the $n$-th harmonic of 
the hadron sector, scaled by a factor 2 and 3 for mesons and baryons, 
respectively. 
Corrections to these linear relations are at least of the order $v_n^2$ and should 
therefore be small.
This gives us a first hint that {\em if} the experimental data exhibit a 
strong $v_4$ component in the particle momentum anisotropy as was 
predicted in Refs.\cite{Kolb03,CKL04} and recently confirmed by the 
STAR collaboration \cite{STAR03}, it is then a true physical component 
in the transverse distribution, and not just an artifact or trivial 
byproduct of the existing large $V_2$ signal. 
We will return to this discussion repeatedly.
We now proceed to study Eqs. (\ref{equ:Vnsolution}) and 
(\ref{equ:Vntildesolution}) under the assumption that higher-order 
coefficients are small corrections in the distribution and can thus 
be neglected from the sums at certain order in $n$. 
%

\section{Anisotropies at midrapidity}
\label{midrapidity}

In the first two years of RHIC physics, a vast amount of data from
Au+Au collisions was collected primarily at midrapidity. Due to 
the symmetry of those systems, the odd Fourier coefficients in 
the expansion of the azimuthal distribution of particle transverse
momentum vanish both on the partonic and hadronic sector.

%
%

{\bf Meson anisotropy at midrapidity:}
Keeping terms up to the order $v_6$ and neglecting higher-order 
corrections, we find from Eq. (\ref{equ:Vnsolution})
that for mesons (up to order $V_6$)
\bea{equ:mesonsaty=0}
V_2 &=& \frac{1}{N} ( 2  v_2 + 2 v_2 v_4 + 2 v_4 v_6) \,, \\ 
V_4 &=& \frac{1}{N} ( 2  v_4 +  v_2^2 + 2 v_2 v_6) \,,\\ 
V_6 &=& \frac{1}{N} ( 2 v_6 + 2 v_2 v_4) \,,
\eea
with $N= 1 + 2(v_2^2 + v_4^2 + v_6^2)$.
Interestingly, we notice that a $V_4$ of the mesons can be generated 
out of the partonic sector even if the partonic distribution 
has only an elliptic component $v_2$, but no $v_4$ contribution. 
This is an effect of the non-linear way how coalescence fuses partons 
into hadrons (see Eq.~(\ref{equ:Fisfsquared})). The resulting $V_4$ is 
however small, namely for a maximum parton anisotropy $v_2$ of about 
10\% as required to describe the experimental hadron anisotropy one 
obtains a $V_4 \sim 1 \%$. 
Neglecting products of 3 or more $v_n$'s in Eq.(\ref{equ:mesonsaty=0}),
we find 
\beq{equ:v4tov2squaredmesons}
\frac{V_4}{V_2^2} \approx \frac{1}{4} + \frac{1}{2} \frac{v_4}{v_2^2}\,.
\eeq
Experimentally \cite{STAR03}, this ratio (measured for all charged hadrons)
is approximately constant as a function of transverse momentum, and has a 
value of about 1.2. Therefore, we can be quite certain that there 
are higher-order harmonics in the Fourier expansion of the azimuthal 
dependence of the parton transverse momentum distribution.
In fact we can deduce from this observation that 
$v_4 (p_T)\sim 2 \, v_2^2(p_T)$. A finite parton $v_4$ is indeed seen 
in the predicted parton anisotropies \cite{CKL04} from the AMPT 
(A Multi-Phase Transport) model \cite{ampt}, although it scales with 
$v_2^2$ with a scaling factor of only about 1. 

%
%
{\bf Baryon anisotropy at midrapidity:}
For the Fourier coefficients of the baryon anisotropy, we again consider 
coefficients up to 6th order. From Eq.~(\ref{equ:Vntildesolution}), we obtain
\bea{equ:baryonsaty=0}
\tilde V_2 =\frac{1}{\tilde N} ( & &3 v_2 + 6  v_2 v_4 + 6 v_4 v_6  +\\
& & + 3 v_2^3 +  3 v_2^2 v_6 + 6 v_2 v_4^2 + 6 v_2 v_6^2 + 3 v_4^2 v_6) 
\,, \nonumber \\ 
\tilde V_4 = \frac{1}{\tilde N} ( & &3 v_4 +3 v_2^2 + 6 v_2 v_6 + \\
&  & + 6 v_2^2 v_4 + 6 v_2 v_4 v_6 + 3 v_4^3 + 6 v_4 v_6^2) \,, \nonumber\\ 
\tilde V_6 = \frac{1}{\tilde N} ( & &3 v_6 + 6 v_2 v_4 + \\ 
& & + v_2^3 + 6 v_2^2 v_6 + 6 v_4^2 v_6 + v_2 v_4^2) \,, \nonumber
\eea
with $\tilde N= 1 + 6 (v_2^2 + v_4^2 + v_6^2) + 6 (v_2^2 v_4 + 2 v_2 v_4 v_6)$.
We note that to obtain the anisotropy values at momentum $p_T$ 
on the left hand side for baryons , we have to use the parton anisotropy 
coefficients at $p_T/3$ on the right hand side.
As already discussed before, there is a linear connection
of quark and baryon anisotropies \cite{MV03},
$\tilde V_n(p_T) \approx 3 \, v_n(p_T/3)$ 
with corrections which are of the order $v_n^2$. 

Again, we express the ratio of the 4-th order coefficient ${\tilde V_4}$
to the square of the elliptic coefficient ${\tilde V_2}$ up to 
second order in partonic $v_n$, and the result is 
\beq{equ:v4tov2squaredbaryons}
\frac{\tilde V_4}{\tilde V_2^2} \approx \frac{1}{3} + \frac{1}{3} 
\frac{v_4}{v_2^2}\,.
\eeq
This confirms the earlier argument, that the value observed by the 
STAR collaboration can only be achieved if there is a significant 
$v_4$ component on the quark sector. 
Together, Eqs. (\ref{equ:v4tov2squaredmesons}) and 
(\ref{equ:v4tov2squaredbaryons}) have the immediate consequence that 
independent of the detailed centrality or momentum dependence of
the quark anisotropies, there is a general relation between baryon 
and meson anisotropies, 
\beq{equ:mesonsandbaryonsratios}
\frac{\tilde V_4}{\tilde V_2^2} (3 \,p_T)
\approx
\frac{2}{3} \, \frac{V_4}{V_2^2} (2 \, p_T)+ \frac{1}{6}.
\eeq
An experimental investigation of the ratios for identified particles 
thus proves promising to further approve or disprove the quark 
coalescence picture, which we leave here as a challenge for future 
experimental analysis and correspondingly for the coalescence model.


\section{Anisotropies for non-zero rapidity or non-identical collision partners}
\label{sec:nonzerorapidity}
At non-zero rapidity the symmetry $x \leftrightarrow -x$, which is 
apparent for collisions of equal nuclei, is broken. 
This allows for the presence of odd Fourier coefficients both in the 
partonic and hadronic sectors. 
This specific symmetry is also lost at midrapidity for collisions of 
non-equal nuclei, such as S+Au. Even if one does not expect to create 
a thermalized system in smaller systems, such as in last years d+Au 
run at RHIC, the following discussion is also useful to quantify and 
relate the different azimuthal coefficients at midrapidity.


{\bf Meson anisotropy:}
We restrict our discussion to the first four coefficients $v_1$ 
to $v_4$ in the partonic distribution. It then follows directly 
from Eq.(\ref{equ:Vnsolution}) that
\bea{equ:mesonsatyneq0}
V_1 &=& \frac{1}{N_y} \left(2 v_1 + 2 v_1 v_2 + 2 v_2 v_3 + 2 v_3 v_4 
\right) \,,\\
V_2 &=& \frac{1}{N_y} \left(2 v_2 + v_1^2 + 2 v_1 v_3 + 2 v_2 v_4 
\right) \,, \\
V_3 &=& \frac{1}{N_y} \left(2 v_3 + 2 v_1 v_2 + 2 v_1 v_4          
\right)  \,, \\
V_4 &=& \frac{1}{N_y} \left(2 v_4 +  v_2^2 + 2 v_1 v_3                        
\right) \,, 
\eea
with $N_y = 1 + 2 (v_1^2 +v_2^2 + v_3^2+v_4^2)$. 

As for the even coefficients in the section before, we see that the 
mesonic $V_n$ is linearly related to the partonic $v_n$ with corrections
that are at least quadratic in $v_n$. 
Even more interesting is that here a meson $V_2$ can be generated 
purely through the partonic directed flow $v_1$ without any dipole
deformation in the initial matter.
Another interesting point is the relation between harmonics of different 
orders. 
As we have already discussed, the experimental data at midrapidity 
indicate a scaling of $V_4$ with $V_2^2$ as also insinuated by the 
coalescence model.
At forward rapidity we can investigate a similar relation for $V_3$. 
We notice that a relation between $V_3$ and $V_1^2$ is unlikely 
because the product of odd harmonics gives rise to an even one, and 
this is indeed the reason why there are no $v_{2n+1}^2$ in $V_{2n+1}$ 
while they are present in $V_{2n}$.  
This suggests a possible relation between $V_3$ and the 
odd-even combination $V_1 V_2$. 
From the above formulas, we find that
\beq{equ:v3tov1v2meson}
\frac{V_3}{V_1 V_2} \approx \frac{1}{2} + \frac{1}{2} \frac{v_3}{v_1 v_2} 
\,\left(+ \frac{1}{2}\frac{v_4}{v_2}\right)\,.
\eeq

The presence of non-negligible odd harmonics at forward rapidity can 
also reflect on the ratio of fourth to second component by adding a new 
term to Eq.(\ref{equ:v4tov2squaredmesons}), i.e., 
\beq{equ:v4tov2meson2}
\frac{V_4}{V_2^2} \approx \frac{1}{4} + \frac{1}{2} \frac{v_4}{v_2^2} + 
 \frac{1}{2}\frac{v_1 v_3}{v_2^2} \,.
\eeq


{\bf Baryon anisotropy:}
The expressions for the baryon anisotropy coefficients are rather 
lengthy but become easily accessible when dropping the terms 
that contain products of three $v_n$ coefficients, i.e., by dropping 
the last two lines of Eq.(\ref{equ:Vntildesolution}). 
The resulting approximate expressions are then
\bea{equ:baryonsatyneq0}
\tilde V_1 &\approx&\frac{1}{\tilde N_y}[3 v_1 + 6 (v_1 v_2 +  v_2 v_3 
+  v_3 v_4)], \\
\tilde V_2 &\approx&\frac{1}{\tilde N_y}[3 v_2 + 3 (v_1^2 + 2 v_1 v_3 
+ 2 v_2 v_4)], \\
\tilde V_3 &\approx&\frac{1}{\tilde N_y}[3 v_3 + 3 ( 2 v_1 v_2 
+ 2 v_1 v_4)], \\
\tilde V_4 &\approx&\frac{1}{\tilde N_y}[3 v_4 + 3 ( 2 v_1 v_3 
+ v_2^2)], 
\eea
and
\bea{equ:Ntildey}
\tilde N_y \approx 1 &+& 6 (v_1^2 + v_2^2 + v_3^2+v_4^2). 
\eea

Again, we notice that the $n$-th coefficient of the baryon spectra is
linearly related to the $n$-th coefficient in the partonic sector, 
and higher-order terms are corrections of the order $v_n^2$. However, 
the above expressions also show that even harmonics in the baryon
spectra can be generated from only odd coefficients in the quark sector, 
even in the absence of even partonic harmonics.

As in the case of mesons, it is of interest to look at the ratio
between the third and the first two components in the azimuthal 
distribution of baryon momentum spectra. 
In the coalescence picture, this is given by
\beq{equ:v3tov1v2baryon}
\frac{\tilde V_3}{\tilde V_1 \tilde V_2} \approx \frac{2}{3} + 
\frac{1}{3} \frac{v_3}{v_1 v_2} 
\,\left(+ \frac{2}{3}\frac{v_4}{v_2}\right)\,.
\eeq
This relation allows us also to investigate the presence of a
non-vanishing $v_3$ at quark level. In fact, if $v_3$ is equal to zero 
the ratio $V_3/V_1 V_2\simeq 0.7$.
Therefore, the existence of a quark $v_3$ component can be
investigated by looking at the magnitude of this ratio from
experimental data, as done for $v_4$. 

Similar to the case of hadrons at midrapidity in collisions of equal mass
nuclei, where baryon ${\tilde V_4}/{\tilde V_2^2}$ and meson 
${V_4}/{V_2^2}$ are related by Eq.(\ref{equ:mesonsandbaryonsratios}), 
we have for the case of non-vanishing odd-order anisotropies
the following relation:
\begin{equation}
\frac{\tilde V_3}{\tilde V_1 \tilde V_2} (3\, p_T) \approx 
\frac{2}{3}\frac{V_3}{V_1 V_2} (2\, p_T) \, + \frac{1}{3}\,,
\end{equation}
if the small $v_4$ contribution is neglected in Eqs.(\ref{equ:v3tov1v2meson}) 
and (\ref{equ:v3tov1v2baryon}).

We want to point out that the above considerations for both mesons and
baryons are applicable in a rapidity region where the hadrons are 
produced by coalescence from the previously established deconfined phase. 
These formulas therefore do not capture the region of highest
rapidity, the fragmentation region, which is located around beam rapidity.

\section{summary}
Using a naive quark coalescence model for hadron production from the
quark-gluon plasma formed in relativistic heavy ion collisions, 
which only allows partons of equal momenta to recombine, we
have expressed the Fourier coefficients of the azimuthal distribution 
of meson and baryon momentum spectra in terms of those of partons.
Neglecting higher-order terms, simple relations are found between 
the ratio of the hadrons 4th order term to the square of their 2nd order
to the corresponding ratio for the partons. Furthermore, a simple 
relation is obtained between such ratios for baryons and mesons,
i.e., Eq. (\ref{equ:mesonsandbaryonsratios}).
This poses an important prediction of the coalescence model
independently of other effects such as flow or jet-quenching.
For hadrons at finite rapidities or spectra from collisions
of non-identical nuclei, odd-order terms are nonzero. We have shown 
that the ratio of their third-order term to the product of their 1st-order 
and 2nd-order terms is also simply related to the corresponding ratio 
for the partons. To account for the observed large ratio ($ \sim 1.2$) 
of hadron 4th-order to the square of 2nd-order anisotropy, a 
larger value ($\sim 2$) for the corresponding ratio for the partonic 
anisotropy is needed. Although this value is larger than that
seen in the AMPT model ($\sim 1$) \cite{CKL04}, which is based on a more 
realistic coalescence picture that includes both the quark momentum 
distributions inside hadrons and also the effect of resonance decays, 
the present schematic study allows one to have a simple understanding 
of the effect of higher-order partonic anisotropies on those of the
observed hadrons. 


\section*{Appendix}
\label{sec:Appendix}

To derive the harmonic coefficients on the hadronic sector, consider 
an arbitrary  distribution function $f(\varphi)$ on the interval 
$\varphi \in [0, 2 \pi]$ expressed in terms of its Fourier series
\beq{equ:fexpansion}
f(\varphi) = \sum_{k=0}^{\infty} f_k \cos k \varphi\,.
\eeq
Squaring this function, as it is required for the meson distributions, 
we obtain
\bea{equ:Fequf2}
F(\varphi) &=& f(\varphi)^2  
= \sum_{k=0}^\infty \sum_{l=0}^\infty f_k f_l \cos k \varphi \cos l 
\varphi  \nonumber \\
&=& \frac{1}{2} \sum_{k=0}^\infty \sum_{l =0}^{\infty} f_k f_l 
\left[ \cos (k\!+\!l) \varphi + \cos (k\!-\!l) \varphi \right]  \,.\nonumber
\eea
From this we then find the term independent of the azimuthal angle
\beq{equ:F0mesons}
F_0 = f_0^2 + \sum_{k=0}^\infty f_k^2 \,,
\eeq
and the amplitude of the $n$th harmonic 
\beq{equ:Fnmesons}
F_n = \frac{1}{2} \sum_{k=0}^n f_k f_{n-k} + \sum_{k=0}^\infty f_k f_{n+k}\,.
\eeq
Substituting $f_0 = 1$ and $f_k = 2 v_k$, we obtain  
the expressions given in Eq.(\ref{equ:Vnsolution}).

For baryons, we have to apply the third power to the function
$f(\varphi)$, which leads to
\bea{equ:Ftildeequf3}
&&\tilde F(\varphi) = f(\varphi)^3 \\ 
&&= \sum_{k=0}^\infty \sum_{l=0}^\infty \sum_{m=0}^\infty
f_k f_l f_m \cos k\varphi \cos l \varphi \cos m \varphi \nonumber \\
&&= \frac{1}{4} \sum_{k, l, m} f_k f_l f_m\left[\cos(k\!+\!l\!+\!m)\varphi 
+ \cos(k\!+\!l\!-m) \varphi \right. \nonumber \\
&&~~ \left.+  \cos (k\!-\!l\! +\!m) \varphi + \cos (k\!-\!l\!-\!m)
\varphi \right] \nonumber \\
&&= \frac{1}{4} \sum_{k, l, m} f_k f_l f_m \left[ \cos(k\!+\!l\!+\!m) 
\varphi + 3 \cos(k\!+\!l\!-\!m) \varphi \right],  \nonumber       
\eea
where the last equality follows from the repeatedly occurring
summation indices in the line above. Again, we can quickly separate 
the terms with a specific amplitude and find
\beq{equ:F0tildebaryons}
\tilde F_0 = \frac{1}{4} f_0^3 + \frac{3}{4} \sum_{k, l}^\infty 
f_k f_l f_{k+l}\,, 
\eeq
and
\bea{equ:Fntildebaryons}
\tilde F_n &= & \frac{3}{4} \sum_{k=0}^\infty \sum_{l=0}^\infty 
f_k f_l f_{n+k+l}  
+\frac{3}{4} \sum_{k=0}^\infty \sum_{l=0}^{n+k} f_k f_l f_{n+k-l} \nonumber \\
&&+  \frac{1}{4} \sum_{k=0}^n \sum_{l=0}^{n-k}f_k f_l f_{n-k-l} \,.
\eea
With the substitution $f_0 = 1$ and $f_k = 2 v_k$, one then obtains
Eq.(\ref{equ:Vntildesolution}).

\section*{Acknowledgments} 
Peter Kolb expresses thanks to the Theoretical Nuclear Physics group 
of Texas A\&M for an invitation which led to this research and 
collaboration. This work was in parts supported by the BMBF and the DFG
(PK), the US National Science Foundation under Grant No. PHY-0098805 
and the Welch Foundation under Grant No. A-1358 (LWC,VG,CMK), as well
as the National Institute of Nuclear Physics (INFN) in Italy (VG)
and the National Science Foundation of China under Grant No. 10105008
(LWC).


\end{narrowtext}
\end{document}